\documentclass[12pt,a4paper]{article}
\usepackage{amsmath,amsfonts,amssymb}
\usepackage{graphicx}
\begin{document}

\parindent=0mm
\parskip 1.5mm

\begin{center}
{\LARGE \bf
  ALEXANDRU PROCA (1897--1955)\\ THE GREAT PHYSICIST}

\bigskip

\small \it

Dorin N. Poenaru

Horia Hulubei National Institute of Physics and Nuclear Engineering
(IFIN-HH),\\ PO Box MG-6, RO-077125 Bucharest-Magurele, Romania
\\ and\\
Frankfurt Institute for Advanced Studies (FIAS), J W Goethe University,
\\Max-von-Laue-Str. 1,  D-60438 Frankfurt am Main, Germany

\end{center}

\begin{abstract}

We commemorate 50 years from A. Proca's death.  Proca equation is a
relativistic wave equation for a massive spin-1 particle. The weak
interaction is transmitted by such kind of vector bosons. Also vector fields
are used to describe spin-1 mesons (e.g. $\rho$ and $\omega$ mesons). After
a brief biography, the paper presents an introduction into relativistic
field theory, including Klein-Gordon, Dirac, and Maxwell fields, allowing to
understand this scientific achievement and some consequences for the theory
of strong interactions as well as for Maxwell-Proca and Einstein-Proca
theories. The modern approach of the nonzero photon mass and the
superluminal radiation field are also mentioned.

\end{abstract}

\section{Introduction and short biography}

A. Proca, one of the greatest physicists of 20th century, was born in
Bucharest on October 16, 1897. Biographical details can be found in a book
editted by his son \cite{pro88b}, where his publications 
are reproduced; see also 
the web site \cite{poew} with many links.
We commemorate 50 years from his death
on December 13, 1955. He passed away in the same year with Einstein.
His accomplishments in theoretical physics are following the developments 
of prominent
physicists including  Maxwell, Einstein and Dirac.
As a young student he started to study and gave public talks on
the Einstein's theory of relativity. 
Then his PhD thesis was dedicated to the Dirac's
relativistic theory of electrons. Proca equation is a relativistic wave
equation
for a massive spin-1 particle. Some of the other relativistic wave eqs. are:
Klein-Gordon eq. describing a massless or massive spin-0 particle; Dirac
equation for a massive spin-1/2 Dirac particle; Maxwell eqs. for a massless
spin-1 particle, etc. In field theory, the Proca action describes a massive
spin-1 field of mass m in Minkowski spacetime. The field involved is a real
vector field. 
Maxwell eqs. and Proca eqs. may be found in many
textbooks \cite{gre00b} 
as important examples of relativistically
invariant formulation of the field equations for a free field; see also 
\cite{itz80b,mor53b}. 
There are many publications mentioning even in the title 
the {\em Einstein-Proca} (e.g. \cite{vui02grg,sci99cqg,tuc97np,der96p}), 
{\em Proca} \cite{bel48pr,ana00fp}, 
or {\em Maxwell-Chern-Simons-Proca} \cite{bel05epj} theory. 

Shortly after finishing in 1915 the {\em Gheorghe Lazar} high school in
Bucharest Proca had to interrupt the studies because the first world war
started. After the war (1918--22) he was student at the Polytechnical School
(PS) specialized in Electromechanics. Next years he was employed by the
Electrical Society, C\^ampina, and in the same time he has also been
assistant professor of Electricity, PS Bucharest.
In 1923 he moved to France because he felt he ``have something to say in
Physics''. In two years brilliantly passed the exams and was graduated by
the Science Faculty, Sorbonne University, Paris. As an experimentalist at
the {\em Institut du Radium} he was very much appreciated by Marie Curie;
she realized he is very interested by the theory and encouraged him to join
the newly founded {\em Institut Henri Poincar\'e}.
In 1930 Proca received the French citizenship and married the
Romanian Marie Berthe Manolesco. He started to work as a {\em Boursier de
Recherches} at his PhD thesis under L. de Broglie's supervision. The thesis
was defended in 1933 in front of a famous commission: Jean Perrin; L.
Brillouin, and L. de Broglie. In 1934 he was one year with E. Schr\"odinger
in Berlin and few months with N. Bohr in Copenhagen, where he met Heisenberg
and Gamow.

From 1936 to 1941 he developed the theory of the massive vector
(spin 1) boson fields governing the weak interaction and the motion of
the spin-1 mesons; {\em Proca equations} refer to this
kind of fields, and are frequently used in Field Theories. Some of the PACS
(Physics and Astronomy Classification Scheme) numbers covered by his work
are: {\em 03. Quantum mechanics, field theories, and special relativity}:
03.50.-z Classical field theories; 03.50.De Classical electromagnetism,
Maxwell equations; 03.70.+k Theory of quantized fields; {\em 04. General
relativity and gravitation}: 04.50.+h Gravity in more than four dimensions,
Kaluza-Klein theory, unified field theories; alternative theories of
gravity; 04.60.-m Quantum gravity; 04.70.-s Physics of black holes; 04.70.Bw
Classical black holes; {\em 11. General theory of fields and particles}:
11.10.-z Field theory; 11.10.Kk Field theories in dimensions other than
four; 11.15.-q Gauge field theories; 11.30.Cp Lorentz and Poincare
invariance; {\em 12. Specific theories and interaction models;
particle systematics}: 12.20.-m Quantum electrodynamics; 12.38.-t Quantum
chromodynamics; 12.40.Vv Vector-meson dominance. Prestigious scientists 
like Yukawa, Wentzel, Taketani, Sakata, Kemmer, Heitler, Fr\"ohlich
and Bhabha, promptly reacted in favour of his equations in 1938. 
W. Pauli \cite{pau45nob,pau41rmp} mentioned Proca's theory in his Nobel
lecture. As a particular sign of his
world-wide recognition one can mention the invitation to attend
in 1939 the {\em Solvay Congress}.  

During the second world war he was for a short time Chief Engineer of the
French Radiobroadcasting Company. Then in 1943 he moved to Portugal where he
gave Lectures at University of Port\^ o. In 1943--45 was in United Kingdom 
invited by the Royal Society and British Admiralty to join the war effort.

After the war in 1946 he started in Paris the {\em Proca Seminar} with many
prestigious invited speakers from France and abroad. This seminar
contributed very much to the education of young French scientistists willing
to work in the field of particle physics. Unfortunately Proca's attempts in 1949 and
1950 to get a chair of Physics at the Sorbonne University and College de
France failed for obscure reasons. Nevertheless in 1950 he accepted to
organize with P. Auger the Theoretical Physics Colloquium of CNRS and in
1951 to be the French delegate at the General Meeting of International Union
of Physics. Starting with 1953 Proca had to fight with a laryngeal cancer
until December 13, 1955 when he passed away.

\section{Particles and fields}

The projection of angular momentum of a particle can be denoted by
$l_z\hbar $, with $l_z$ an integer $-l \leq l_z
\leq l$. The intrinsic angular momentum, $s$, is called spin. According to
the spin values the particles obey to the Fermi-Dirac or Bose-Einstein
statistics. A fermion ($s=(n+1/2)\hbar $, $n$ integer) is called spinor if 
$s=\hbar/2$, e.g. leptons ($e, \nu, \mu, \tau$) or spinor-vector 
if $s=(3/2)\hbar$. A boson ($s=n\hbar$) can be scalar ($s=0$), vector
($s=\hbar$, e.g. $\omega$, $\rho$ mesons, 
$m=0$ photons, weak massive $W^{\pm}, Z^0$ bosons,
$m=0$ gluons), or tensor ($s=2\hbar$, e.g. graviton).
{\em Vector bosons play a central role as they
are mediators of three (of the four) fundamental interactions: 
electromagnetic, weak, and strong. Tensor
bosons are assumed to mediate the gravitation} (see the Table \ref{tab1}). 
\begin{center}
\begin{table}
\caption{The four fundamental interactions.}
\bigskip
\begin{center}
\begin{tabular}{|p{3.3 cm}|p{2.1 cm}|p{5.2 cm}|} \hline
FORCE       	& RANGE	&TRANSMITTED BY BOSONS \\  \hline

gravity & long & graviton, massless, ~~~~spin 2 \\ \hline
electro\-mag\-ne\-tism & long & photon ($\gamma$),
massless, ~spin 1 \\ \hline
weak interaction & short & $W^{\pm}$,
$Z^0$, heavy, ~~~~~~~~spin 1\\ \hline
strong interaction& short & gluons
($g$), massless, ~~spin 1\\ \hline

\end{tabular} \label{tab1}
\end{center}
\end{table}
\end{center}
A particle is a localized entity. A field is an assignment of
a quantity to every point of space. Einstein's special theory of relativity
allows for the existence of scalar, vector, tensor, fields.  Examples of 
relativistic fields: Klein-Gordon
($s=0$); Dirac ($s=1/2$); Proca ($s=1$, $m\neq 0$); Maxwell ($s=1$, $m=0$); 
Rarita-Schwinger ($s=3/2$), and Gravitation ($s=2$).

In particle physics, quantum field theories, cosmology and quantum gravity,
the {\em  natural units, $\hbar=c=1$,}  are used. The units of length, time
and mass are expressed in GeV: 1~meter$=5.07\times 10^{15}$~GeV$^{-1}$;
1~second$=1.52\times 10^{24}$~GeV$^{-1}$ and 1~kg$=5.61\times 10^{26}$~GeV. The Newton's
gravitational constant is given in terms of Plank's mass $G=M^{-2}_{Pl}$,
where $M_{Pl}=1.22\times 10^{19}$~GeV. In particle physics gravity becomes
important when energies or masses approach $M_{Pl}$.

{\em Geometrical units, $c=G=1$,}  are used in classical general relativity,
and every quantity is expressed in units of length, e.g. $\hbar=L^2_{Pl}$
where the Planck length $L_{Pl}=1.6\times 10^{-33}$~cm. In gravitation
(general relativity) quantum effects become important at lenghth scales
approaching $L_{Pl}$. We shall use in the following the natural units.

A contravariant and covariant four vector has four components
\begin{equation}
a^\mu=(a^0;a^1,a^2,a^3) \; ; \; \; \; \; a_\mu=(a_0;a_1,a_2,a_3)
\end{equation}
e.g $x^\mu=(t;x,y,z)=(x^0;x^1,x^2,x^3)$, $p^\mu=(E;\vec{p} \;)$.

The metric tensor (covariant components) 
\begin{equation}
g_{\mu \nu} = \left ( \begin{array}{cccc}
                                          1 &  &  & \\
                                            &-1&  & \\
                                            &  &-1& \\
                                            &  &  &-1\\
                                           \end{array}   \right )
\end{equation}
For the Lorentz metric the contravariant form $g^{\mu \nu}=g_{\mu \nu}$. One
can change a contravariant 4-vector into a covariant one by using the
relationship $a_\mu=g_{\mu \nu}a^\nu$. One assumes the
Einstein's convention: summ over repetead indices. 

The scalar product
\begin{equation}
a_\mu b^\mu=a^0b^0 -a^1b^1 -a^2b^2 -a^3b^3=a^0b^0 -\vec{a}\vec{b}
\end{equation}
The derivatives  
\begin{equation}
\partial ^\mu = \frac{\partial}{\partial x^\mu}=\left (
\frac{\partial}{\partial t}; -\frac{\partial}{\partial x},
-\frac{\partial}{\partial y}, -\frac{\partial}{\partial z} \right
)=(\partial ^0; -\nabla) ; \; \; \partial_\mu=\frac{\partial}{\partial x^\mu}
= (\partial ^0; \; \nabla)
\end{equation}
\begin{equation}
\partial_\mu a^\mu = \frac{\partial a^0}{\partial t} + \nabla \vec{a}
\end{equation}
\begin{equation}
\partial_\mu \partial^\mu = \frac{\partial^2}{\partial t^2} -
\frac{\partial^2}{\partial x^2} - \frac{\partial^2}{\partial y^2} -
\frac{\partial^2}{\partial z^2} = \partial_0^2 - \nabla ^2 = \partial^\mu
\partial_\mu = \Box
\end{equation}
$\nabla ^2$ is the Laplacian and $\Box= \left (  
\frac{\partial^2}{\partial t^2} - \nabla ^2\right )$ is the d'Alembertian.

Volume elements $d^4x=d^3xdt=d^3xdx^0 \; ; \; \; d^3x=dxdydz=dx^1dx^2dx^3$.

For a conservative force, $\vec{F}$, 
there is a potential $V$ such that 
$\vec{F}=-\nabla V$. 
The Lagrangian in terms of the generalized coordinates $q_i$ and
velocities $\dot{q_i}$ is given by $L=T-V$, where $T$ is the kinetic energy.
According to the Hamilton's principle, the classical action developed during
the time interval $t_2 - t_1$
\begin{equation}
S=\int_{t_1}^{t_2}L(q_i, \dot{q_i}, t) dt
\end{equation}
has a stationary value ($\delta S=0$) for the dynamical path of motion.
The Euler-Lagrange eq. 
\begin{equation}
\frac{d}{dt}\left(\frac{\partial L}{\partial \dot{q_i}} \right) -
\frac{\partial L}{\partial q_i} =0
\end{equation}
is obtained by performing a calculus of variations 
which leads to the Newton's second law in classical mechanics of a
point particle.

In both special and general relativity one seeks covariant eqs. in which
space and time are given equal status. The above defined action is not
covariant. In field theory we replace the $q_i$ by a field $\varphi (x)$
where $x\equiv (t,\vec{x})$. A covariant form of action involves 
the Lagrangian density, ${\cal{L}}={\cal{L}}(\varphi , \partial_\mu \varphi)$,
which is a functional integrated over all space-time 
\begin{equation}
{\cal{S}}[\varphi_i]=\int {\cal{L}}[\varphi_i , \partial_\mu \varphi] \; d^4x
\end{equation}
The Lagrangian is the spatial integral of the density.

The least action principle leads to the Euler-Lagrange equation
\begin{equation}
\frac{\delta}{\delta \varphi}{\cal{S}}=-\partial_\mu \left(\frac{\partial
{\cal{L}}}{\partial(\partial_\mu\varphi)}\right) +
\frac{\partial{\cal{L}}}{\partial \varphi}=0
\end{equation}
The covariant momentum density is defined by 
\begin{equation}
\Pi^\mu=\frac{\partial {\cal{L}}}{\partial(\partial _\mu\varphi)}
\end{equation}
The energy-momentum tensor
\begin{equation}
T_{\mu \nu} \equiv \pi_\mu\partial_\nu\varphi - g_{\mu \nu}{\cal{L}}
\end{equation}
is analogous to the definition of the point particle Hamiltonian.

The quantum mechanics can describe a system with a fixed number of particles
in terms of a many-body wave function. The relativistic quantum field theory
with creation and annihilation operators was developed in order to include
processes (e.g. $n \rightarrow p + e + \bar{\nu _e}$ or $e^+e^- \rightarrow
2\gamma$) in which the number of particles is not conserved, and to describe
the conversion of mass into energy and vice versa.

\subsection{Klein-Gordon field }

For a massive ($m\neq 0$) scalar {(spin $0$)} and neutral (charge zero)
field, the Lagrangian density (function of the fields
$\phi$ and their $x, y, z, t$ derivatives) is 
\begin{equation}
{\cal{L}}=(1/2)[(\partial _\mu \phi)(\partial ^\mu \phi) - m^2\phi ^2]
\end{equation}
The Euler-Lagrange formula requires 
\begin{equation}
(\Box + m^2)\phi=0
\end{equation}
i.e. the Klein-Gordon equation.
It was quantized by Pauli and Weisskopf in 1934. 
Relativistic wave equations are invariant under Lorentz transformations,
expressing the invariance of the element of 4-vector length,
$ds^2=dt^2-(dx^2+dy^2+dz^2)$.
The Klein-Gordon equation was historically rejected as a fundamental quantum
equation because it predicted negative probability density.

For free particles with an internal degree of freedom (e.g. electric charge)
the real valued field is replaced by complex fields $\phi ^* \neq \phi$,
hence nonhermitean field operators $\hat{\phi}^\dagger \neq \hat{\phi}$. 

\subsection{Dirac field }

Dirac was looking for an equation linear in $E$ or in
$\partial /\partial t$.
For a massive spinor (spin $1/2$) field the Lagrangian density is 
\begin{equation}
{\cal{L}}=\bar{\psi}(i\gamma ^\mu \partial _\mu - m)\psi
\; \; ; \; \; \; \bar{\psi}=\psi ^*\gamma ^0 \; \; Dirac \; \; adjoint
\end{equation}
where the four $4\times4$ 
Dirac matrices $\gamma ^\mu$ ($\mu=0, 1, 2, 3$) satisfy the Clifford
algebra 
\begin{equation}
\{\gamma ^\mu, \gamma ^\nu \}=
\gamma ^\mu \gamma ^\nu + \gamma ^\nu \gamma ^\mu = 2g^ {\mu \nu}
\end{equation}
The corresponding equation of motion 
is the Dirac equation
\begin{equation}
(i\gamma ^\mu \partial_\mu - m)\psi=0 \; \; ; \; \; \; i(\partial_\mu
\bar{\psi})\gamma ^\mu + m\bar{\psi}=0
\end{equation}
One set of  $4\times4$ matrices is  
\begin{equation}
\gamma ^0 =\left (\begin{array}{cc} I & 0\\0 & I\\  \end{array}
\right ) , \; \; \; \; \gamma ^i =\left (\begin{array}{cc} 0 & \sigma ^i\\
-\sigma ^i & 0\\ \end{array} \right )
\end{equation}
where one has $2\times 2$ matrices: identity $I$, zero and 
Pauli matrices:
\begin{equation}
\sigma ^1=\left (\begin{array}{cc} 0 & 1\\1&0\\ \end{array} \right ) , \;
\; \sigma ^2=\left (\begin{array}{cc} 0 &-i\\i&0\\ \end{array} \right ) , \;
\; \sigma ^3=\left (\begin{array}{cc} 1&0\\0&-1\\ \end{array} \right )
\end{equation}
Quantization of the Dirac field is achieved by replacing the spinors by
field operators and using the Jordan and Wigner quantization rules.
Heisenberg's eq. of motion for the field operator $\hat{\psi}(\vec{x}, t)$
reads 
\begin{equation}
i\frac{\partial}{\partial t}\hat{\psi}(\vec{x},t)=[\hat{\psi}
(\vec{x},t),\hat{H}]
\end{equation}
There are both positive and negative eigenvalues in the energy spectrum.
The later are problematic in view of
Einstein's energy of a particle at rest $E=mc^2$. Dirac's way out of the
negative energy catastrophe was to postulate a Fermi sea of antiparticles.
This genial assumption was 
not taken seriously until the positron was discovered in 1932 by
Anderson.

\subsection{Maxwell field }

In classical field theory the differential form of Maxwell eqs. are
given by Gauss's, Amp\`ere's and Faraday's laws  plus Maxwell's extensions. For
homogeneous materials:
\begin{equation}
\nabla \cdot \vec{D} = \rho \; ; \; \; \; \nabla \cdot \vec{B} =0 \; ; \;
\; \; \vec{D} = \varepsilon \vec{E}\; ; \; \; \; \vec{B} = \mu \vec{H}
\end{equation}
\begin{equation}
\nabla \times \vec{E} = -\frac{\partial \vec{B}}{\partial t}
\end{equation}
\begin{equation}
\nabla \times \vec{B}/\mu = \vec{j} + \varepsilon
\frac{\partial \vec{E}}{\partial t}
\end{equation}
Photons are assumed to be massless (forces of infinite range).
Electrodynamics is gauge invariant ($\varphi$ and $\vec{A}$ are not unique): 
Maxwell eqs. do not change under gauge
transformation (with $\chi =\chi (\vec{r},t)$ an arbitrary differentiable function)
\begin{equation}
\vec{A} \rightarrow \vec{A} + \nabla \chi (\vec{r},t) \; ; \; \; \;
\varphi \rightarrow \varphi -\frac{\partial \chi}{\partial t}
\end{equation}
The 4-potential and current density  are
\begin{equation}
A^\mu=(\varphi;\vec{A} \; )=(A^0;\vec{A} \; ), \; \; j^\mu = (\rho; \vec{j}
\; )
\end{equation}
Scalar, $\varphi$, and vector, $\vec{A}$, potentials are introduced via 
\begin{equation}
\vec{E}=-\nabla \varphi -\partial
\vec{A}/\partial t, \; \; \vec{B}=\nabla
\times \vec{A}
\end{equation}
The four vector potential 
\begin{equation}
A^\mu =(\varphi, \vec{A})\; ; \; \; \; \;
g_{\mu \nu }A^\mu A^\nu =\varphi ^2-\vec{A}^2
\end{equation}
The antisymmetric field-strength tensor 
\begin{equation}
F^{\mu \nu} = \partial ^\mu A^\nu
-\partial ^\nu A^\mu
\end{equation}
with components $F^{0i}=\partial ^0 A^i - \partial ^i
A^0 = - E^i$ and $F^{ij} = \partial ^i A^j - \partial ^j A^i = - \epsilon
^{ijk}B^k$. The Levi-Civita symbol $\epsilon ^{ijk}$ is antisymmetric under
exchange of any two indices. 

The Lagrangian density is
\begin{equation}
{\cal{L}} = (E^2 -B^2)/2 -\rho V + \vec{j}\vec{A}=
 -\frac{1}{4}F_{\mu \nu}F^{\mu \nu} - j_\mu A^\mu
\end{equation}
The corresponding Euler-Lagrange eqs. are Maxwell's eqs.
Kaluza and Klein attempted to unify the gravitation and electromagnetic
theories by extending general relativity in 5 dimensions.

\subsection{Proca (massive vector boson) field }

Proca extended the Maxwell eqs. in quantum field theory. In 1934 for massive
positively charged particles with spin there were two alternatives:
Dirac eqs. 
having a spectrum of positive and negative
energies, a positive charge and a finite spin
($\pm E, +q, s\neq 0$) or Pauli-Weisskopf 
based on Klein-Gordon eqs. ($\pm E, \pm q, s=0$). Proca worked out new
equations which would allow for  positive and negative 
energies, both signs of the
charge, and a finite spin ($\pm E, \pm q, s\neq 0$).
For a massive vector boson (spin~1) field the Proca equation \cite{pro36jpr} 
\begin{equation}
\Box A^\nu - \partial ^\nu(\partial_\mu A^\mu) + m^2 A^\nu = j^\nu
\end{equation}
is obtained as a Euler-Lagrange eq. emerging from the Lagrangian
\begin{equation}
{\cal{L}}=-\frac{1}{4}F_{\mu \nu}F^{\mu \nu} + \frac{1}{2}m^2A_\mu A^\mu
-j_\mu A^\mu
\end{equation}
after expressing the field-strength tensor, $F^{\mu \nu}$,
in terms of the four potential $A^\mu$. 
The Maxwell field is a massless ($m=0$) Proca field.
In contrast to the Maxwell field the Lorentz condition is fulfilled by Proca
field. 
In his Nobel lecture  \cite{pau45nob}, W. Pauli noted:
``The simplest cases of one-valued fields are the scalar field and a field
consisting of a four-vector and an antysimmetric tensor like the potentials
and field strengths in Maxwell's theory. While the scalar field is simply
fulfilling the usual wave equation of the second order in which the term
proportional to $\mu^2$ has to be included, the other field has to fulfill
equations due to Proca which are generalization of Maxwell's equations
which become in the particular case $\mu=0$.''

Assuming $m\neq 0$ one has $\partial_\nu A^\mu
= (1/m^2)\partial_\nu j^\nu$. 
If the source current is conserved ($\partial_\nu j^\nu =0$) or if there 
are no surces ($j^\nu =0$) it follows that $\partial_\nu A^\nu =0$. 
The field eq. gets simplified $(\Box + m^2)A^\nu=0$ for free particles,
leading to four Klein-Gordon eqs. for projections. 

After Yukawa's \cite{yuk35pmsj} hypothesis of a particle mediating the
nuclear interaction this particle was initially called mesotron, or
alternatively the Proca particle (see for example the ref.~\cite{bat41pr}).

\section{Yukawa and the strong interaction (color force)}

Hideki Yukawa (1907-1981) was born in Tokyo as a third son of Takuji Ogawa.
Genyo Yukawa adopted\footnote[1]{Adoption was a common practice in Japan 
in a family without son.} him when he married Sumiko Yukawa in 1932. Hideki
won the
Nobel Prize in 1949 {\em for his prediction (in 1934) of the existence of
mesons on the basis of theoretical work on nuclear forces} \cite{yuk35pmsj}. 
His potential (of a Debye type) 
\begin{equation}
g^2\frac{e^{-\lambda r}}{r}
\end{equation}
can explain the short range of the strong interaction. At that time
there was no link between the quantum theory of fields and nuclear theory,
except the Fermi's $\beta$-decay theory. 
When he was only 27 years old he predicted the existence of new particles
now called pions.
Yukawa calculated a mass $\sim 200m_e$ ($m_e$ is the electron mass). 
By analogy with photons mediating the
electromagnetic interaction he assumed the nuclear forces, acting between nucleons,
are mediated by such bosons. As he suggested, the study of cosmic rays gave
the first experimental evidence of the new particles.  
Yukawa employed a scalar field equation. The right vector field was
introduced by Proca (see e.g. \cite{pau41rmp}: ``This case holds the
center of current interest since Yukawa supposed the meson to have the spin
1 in order to explain the spin dependence of the force between proton and
neutron. The theory for this case has been given by Proca''). At present we
know that the strong interaction is mediated by gluons.

Yukawa was ahead of his time and found the key to the problem of nuclear
forces. His paper was unnoticed until 1937 when Anderson 
announced his dicovery in cosmic rays of a particle with  a mass 
similar to that
required by Yukawa's theory. Soon it was clear that Anderson's ``mesotron''
(now the muon or $\mu$-meson) did not possess the right properties;
it is in fact a lepton.
Cecil Powell (Nobel prize in 1950) discovered in 1947 the $\pi$-mesons
(pions) in cosmic rays.

The name meson means middle-weight between electron and nucleon. Cosmic
rays contained two intermediate mass particles: muon and pion.
The muon is a lepton (a heavy counterpart to the electron) and
not a meson (although it is still called  $\mu$-meson for historical reasons), 
but the pion was a true meson of the kind predicted by Yukawa.
Around 1970 there were many theories \cite{gre70s} attempting to explain the
nuclear interaction.
Presently according to quantum chromodynamics, the strong interaction,
mediated by massless gluons, affects only quarks and antiquarks; it binds
quarks to form hadrons (including proton and neutron). 
There are 8 types of gluons.

{\em
The theory of the massive vector bosons with spin 1 was developed by A. Proca. 
Such bosons ($W^\pm$ and $Z^0$ bosons) are mediators of the weak 
interaction. Proca's equations are also used to describe spin 1 mesons, 
e.g. $\rho$ and $\omega$ mesons.}

\section{Nonzero photon mass and the superluminal \\radiation field}

The effects of a nonzero photon rest mass  can be incorporated into
electromagnetism through the Proca eqs.
The massive electromagnetic field is described by {Maxwell-Proca 
eqs.}
\begin{equation}
\nabla \cdot \vec{E} =\frac{\rho}{\varepsilon_0} - \mu_\gamma^2 \varphi
\; ; \; \; \; \nabla \cdot \vec{B} =0
\end{equation}
\begin{equation}
\nabla \times \vec{E} = -\frac{\partial \vec{B}}{\partial t} \; ; \; \; \;
\nabla \times \vec{B}=\mu_0 \vec{j} + \mu_0\varepsilon_0\frac{\partial
\vec{E}}{\partial t} -\mu_\gamma^2\vec{A}
\end{equation}
where $\mu_\gamma^{-1}=\hbar/(m_\gamma c)$ is the 
Compton wave-length of a photon with mass $m_\gamma$.
Implications of a massive photon:
variation of $c$; longitudinal electromagnetic 
radiation and gravitational deflection;
possibility of charged black holes; the existence of magnetic monopoles
\cite{lak04pla}; 
modification of the standard model \cite{dva05prl}, etc.
An upper limit for the photon rest mass \cite{tu05rpp} 
is {$m_\gamma\leq 1\times 10^{-49}$~g$\equiv 6\times 10^{-17}$~eV.}

The concept of a nonzero rest mass graviton may be defined \cite{arg97aj} in
two ways: phenomenologically, by introducing of a mass term in the
linear Lagrangian density, as in Proca electrodynamics, and 
self-consistently, by solving Einstein's equations in the conformally flat
case. The rest mass of the graviton was given in terms of
the three fundamental constants: gravitational, Planck, and light velocity.
The Einstein-Proca equations, describing a spin-1 massive vector
field in general relativity, have been studied \cite{vui02grg}
in the static spherically-symmetric case. 
It was shown \cite{tou99t} that a special case of the metric-affine
gauge theory of gravity is effectively equivalent to the coupled
Einstein-Proca theory.

Einstein-Proca field eqs. are frequently discussed in connections with dark
matter gravitational interactions \cite{bei04ijtp}. 
At the level of string
theories there are hints that non-Riemannian models, such as
{Einstein-Proca-Weyl} theories \cite{sci99cqg} 
may be used to account for the dark matter.
Other developments: {Proca-Wightman} field \cite{yao75cmp};
Maxwell-Chern-Simons-Proca model~\cite{baz03jpa}.

Superluminal (faster than light) particles, {\em tachyons}, with an
imaginary mass of the order of $m_e/238$, can be described by a real Proca
field with a negative mass square \cite{tom01p}. They could
be generated in storage rings, jovian magnetosphere, and supernova remnants.

In conclusion, A. Proca lived in a period of great discoveries and
development of quantum field theories to which he contributed in an
essential way. After about eighty years of use, Proca equation of the vector
boson fields remains one of the basic relativistic wave equation. The weak
interaction is transmitted by such kind of vector bosons. Also vector fields
are used to describe spin-1 mesons such as $\rho$ and $\omega$ mesons.

{\bf Acknowledgement} The author is grateful to Prof. M. Vi\c{s}inescu for
critical reading of the manuscript.

\end{document}